\documentclass[12pt,a4paper]{article}

\usepackage{subfigure}
\usepackage{graphicx}
\usepackage{epsfig}
\date{ }
\begin{document}
\title {Characterization of Jets in Relativistic Heavy Ion Collisions} 

\author {S. C. Phatak{\footnote {email:  
phatak@iopb.res.in} and P. K. Sahu{\footnote {email:
  pradip@iopb.res.in}} } \\ {\it Institute of Physics,
 Bhubaneswar 751 005, India} }
\maketitle

\begin{abstract}
Jet quenching is considered to be one of the signatures of the
formation of quark gluon plasma. In order to investigate the jet 
quenching, it is necessary to detect jets produced in relativistic 
heavy ion collisions, determine their properties and compare those with 
the jets one obtains in hadron-hadron or $e^+-e^-$ collisions. In this 
work, we propose that calculation of flow parameters may be used to 
detect and characterize jets in relativistic heavy ion collisions.
\end{abstract}

\section{Introduction}

The quenching of jets in relativistic heavy ion collisions has been
proposed to be one of the signatures of the formation of quark gluon
plasma (QGP)\cite{jet-quench}-\cite{quench1}. For a jet produced in heavy ion 
collision, the leading energetic parton of the jet will
have to traverse through a medium consisting of QGP before it fragments 
and a jet is produced. During its passage through the plasma, 
it may loose some of its energy because of its interaction with the
constituents of the plasma. It would also produce quarks and gluons
during this period\cite{jet-quench}-\cite{quench2}. This effect seems to 
have been
observed in the recent experiments\cite{expt1,expt2}, although the
observation is somewhat indirect. The quenching of the jet is expected 
to have several consequences on the characteristics of the jet. For 
example, one expects that the energy carried by the hadrons constituting
the jet should decrease due to jet quenching. This is because the 
leading parton of the jet looses its energy via gluon radiation
while traversing through QGP and that results in the decrease in the 
energy carried by the hadrons in the jet. Another possible consequence
of jet quenching is, the number of hadrons produced in the  
jet might increase because secondary quarks and gluons that are produced
during the passage of the leading parton through the plasma. 
One more consequence of the interaction of the
leading parton in the plasma is that there would be an increase in the
opening angle of the jet. This would lead to a larger spread of the
jet particles in azimuthal angle as well as rapidity. 

The properties of jets produced in $e^+-e^-$ and hadron-hadron collisions
have been well investigated\cite{qcdjets1,qcdjets2} and the systematics 
of number of particles in a jet and the opening angle of the jet have 
been done. There is also a good
theoretical understanding of these in terms of perturbative QCD and
jet fragmentation functions\cite{qcdjets2,qcdjets3}. In case of
nucleus-nucleus collisions, on the other hand, the identification and 
characterization of jets are very difficult because during the collision
process, a large number of hadrons are produced and the 
particles belonging to the jet constitute a very small fraction of
these. Therefore, one cannot easily isolate the particles produced in
a jet from the background particles
and determine the properties of jets in a simple fashion. 
%Added
The jet characterization is made further difficult by small 
probability of jet production. Assuming a conservative value
of 100 nb as the hard jet production cross section\cite{qcdjets2}
in nucleon-nucleon collision, the probability of jet production in
central gold-gold collision is about $10^{-3}$.
%Added
Indeed the particles produced in a jet have large momenta and they are 
emitted in a narrow cone in $\eta$ ( the rapidity ) and $\phi$ ( the azimuthal 
angle ) and the background particles are distributed over a large range
of $\eta$ and spread more or less uniformly in $\phi$. But there is no 
obvious method of removing the background particles to extract the 
properties of the jet. Typically, the dispersion in the rapidity and
azimuthal angle of the jet particles is quite small ( $\Delta =
\sqrt{\Delta_\eta^2 + \Delta_\phi^2}$ with $\Delta_\eta \sim
\Delta_\phi \le 0.5$ ). It would be useful to characterize the jets
produced in heavy ion collisions in terms of the quantity $\Delta$
defined above as well as in terms of the number of particles in the
jet and the energy carried by the jet particles. This will give a
direct measure of jet quenching and jet fragmentation function.
Further, measurement of these quantities as a function of impact 
parameter ( which is related to the total multiplicity or transverse
energy in an event ) would also be of interest. This is because the
amount of quark matter traversed by the leading parton in the jet, and 
therefore the amount of jet quenching is expected to depend on the 
impact parameter. Further, in case of large impact parameter collisions, 
one would expect that the quenching would be different if the jet 
direction is in  or out of the reaction plane. 

Methods, such as tagging the jet with photons\cite{tag-phot} or
di-leptons\cite{tag-phot,tag-dielept} have been proposed to detect the
jets and determine their properties in relativistic heavy ion collisions.
These methods are indirect and they do not give full information on the
jets. In this note, we have proposed a method of locating and
identifying a jet from the data on hadron momentum distributions for a
given event. The method uses the flow parameters\cite{flow} for this 
purpose. Our analysis indicates that the method is capable of estimating 
the number of particles in the jet as well as the jet opening angle 
reasonably well. 
We show that with the analysis of flow parameters for different  
transverse energy cuts, it is possible  to estimate the number of
particles in a jet and the opening angle of the jet to some extent.

Investigation of flow parameters in heavy ion collisions is being done 
for quite some time\cite{flow1}. The main reason of this analysis is to 
determine the collective motion of the particles produced in the 
collisions. The focus here is on directed and elliptical flows which are 
expected to shed some light on the equation of state of the hot and 
dense matter. The flow introduced by jets is not exactly a collective 
effect.  So, here we are advocating the measurement of flow for very
different physical effect. 

In the following, we describe the method in Section II, apply the
method to synthetic data as well as simulated data\cite{simul} in
Section III and describe the results and conclude in Section IV.

\section{The Method}

Given the transverse momentum distribution of particles produced in an 
event of relativistic heavy ion collision, one can compute the 
quantities  $b_m(p_0)$ for all the  particles having transverse 
momentum larger than some transverse momentum $p_0$;
\begin{eqnarray}
  \label{eq:1}
  b_m(p_0) & = &\frac{\sum_{j} \cos(m \phi(j) )}{\sum_{j}} \nonumber \\
  & = & \frac{1}{N(p_0)} \sum_{j, p_T(j) > p_0} \cos(m \phi(j) ).
\end{eqnarray}
Here $\phi(j)$ is the azimuthal angle of $j^{th}$ particle, the 
summation index $j$ runs over all particles having
transverse momentum larger than $p_0$ and $N(p_0)$ is
the number of particles having transverse momentum larger than
$p_0$. The coefficients $b_m$  are related to the flow coefficients ( 
see later ) and for different values of $m$, these have specific 
meaning. For example, $b_1$ and $b_2$ are related to the directed and 
elliptic flow respectively. Generally, the flow 
coefficients for $m$ larger than 2 are not considered useful. Here, 
however, we shall be arguing that these would be needed to get 
information about jets. The coefficients $b_m$'s are also small and 
difficult to determine for $m > 2 $ and, therefore not emphasized in the
literature. However, recently it has been argued that, the 
fourth order flow coefficient is important in understanding of the 
physics of extreme energy loss and getting the information on dynamical 
evolution of the system{\cite{v4}. The reasons for introducing momentum 
cut will be explained later.

Theoretically the flow coefficients $v_m$'s are measured with respect to 
the reaction plane.  Their definition is:
\begin{eqnarray}
  v_m = \Big < \cos(m(\phi_i - \phi_E))\Big >,
\end{eqnarray}
where $\phi_E$ defines the reaction plane and $\Big < \cdots \Big >$ 
implies averaging of the quantity $\cdots$ over all particles. Since, 
in an experiment, it 
is not possible to define the reaction plane, the flow coefficients
are obtained by the pairwise correlation of all the particles.
It is easy to show that the square of the flow coefficients, $v^2_m$ 
are given by
\begin{eqnarray}
  v^2_m & = & \Big < \cos(m(\phi_i - \phi_j))\Big >_{i,j} \nonumber \\
 & = & \frac{\sum_{i, j} \cos (m (\phi_i - \phi_j) )}{N^2(p_0)} .
\label{eq-av}
\end{eqnarray}
We shall now consider the flow coefficients $v_m(p_0)$'s for further 
discussions. If the particles produced in the reaction are uniformly 
distributed in azimuthal angles, theoretically all $v_m(p_0)$'s are 
zero. In case of the experimental data
$v_m^{p_0}$ approaches zero as the  number of particles goes to infinity.
Thus, because of statistical fluctuations, the experimentally 
determined values of $v_m(p_0)$'s are always nonzero. On the other
hand, if all the particles have same azimuthal angle  
$\phi$, all $v_m$'s will be unity.

It turns out that the computation of pairwise correlated quantities 
$v_m^2$'s of eq(\ref{eq-av}) not only removes the uncertainty in the 
knowledge of the reaction plane but helps in reducing the statistical 
fluctuations also. 

To begin with, let us consider that the particles are distributed 
according to an azimuthal probability distribution function $P(\phi)$. The
flow coefficients are then defined as
\pagebreak
\begin{eqnarray}
v_m^2 & =& <\cos(m (\phi_1-\phi_2))> \nonumber \\
& =& \frac{\int d\phi_1 d\phi_2 \cos(m (\phi_1 - \phi_2)) 
P(\phi_1)P(\phi_2)}
	{\int d\phi_1 d\phi_2 P(\phi_1) P(\phi_2)}\nonumber \\
       &=& \frac{\int d\phi_1 \int d\phi_2 [\cos(m\phi_1)\cos(m\phi_2)+ 
        \sin(m\phi_1)\sin(m\phi_2)]
         P(\phi_1)  P(\phi_2)}{\int d\phi_1 P(\phi_1) 
       \int d\phi_2 P(\phi_2)}.
\label{eq-vm}
\end{eqnarray}

\subsection{One Jet}

Let us now consider an event in which a jet with $N_J$ number of 
particles are produced. In addition, let us assume that the event 
consists of $N_b$ non-jet ( background ) particles and the transverse 
momentum and the azimuthal angle of these particles is known. Let us 
further assume that the background particles are distributed uniformly 
in the azimuthal angle $\phi$ and the azimuthal angles of the jet 
particles are uniformly distributed within the azimuthal angle $\phi_0 
- \frac{\Delta \phi}{2}$ and $\phi_0 +  \frac{\Delta \phi}{2}$. We shall
also assume that the particles in a specified rapidity window are chosen
for the analysis. For such a case, the probability distribution function 
would be
\begin{eqnarray}
P(\phi) & = & P_1(\phi) + P_2(\phi), 
\label{dist_fun}
\end{eqnarray}
where
\begin{eqnarray}
P_1(\phi) & = &  \frac{N_b}{2 \pi N} ~~~~~~~~ 0 < \phi < 2 \pi \nonumber \\
P_2(\phi) & = &  \frac{N_J}{\Delta \phi N} ~~~~~~~~
 \phi_0-\frac{\Delta \phi}{2} < \phi <  \phi_0+\frac{\Delta \phi}{2} ,
\nonumber \\
& = & 0  ~~~~~~~~~~~~~~~~~~ {\rm otherwise} \nonumber
\end{eqnarray}
and $N = N_b + N_J$. 
For the probability distribution given above, the flow coefficients 
$v_m^2$, as defined in eq(\ref{eq-vm}) can be computed analytically. For 
this, one may choose the x-axis along $\phi_0$. Further, all the 
integrals in the numerator of eq(\ref{eq-vm}) vanish for $P_1(\phi)$ and
we get
\begin{eqnarray}
v_m^2(p_0) 
&=&\frac{N_J^2}{N^2}\Big [\frac{\sin(\frac{m\Delta\phi}{2})}
{(\frac{m\Delta\phi}{2})}\Big ]^2 
=\frac{N_J^2}{N^2}\Big [ j_0 \Big ( \frac {m \Delta \phi} {2} \Big ) 
\Big ]^2.
\label{flow}
\end{eqnarray}
The $\sin$ integral terms in eq(\ref{eq-vm}) vanished due to symmetry 
distribution function $P_1$.
If the opening angle $\Delta \phi$ is
small enough, the Bessel function $j_0$ can be expanded in power
series. Keeping first two terms we get,
\begin{eqnarray}
v_m^2 & \sim & \frac{N_J^2}{N^2} \Big [ 1
- \frac{( \Delta \phi)^2 m^2}{12} + O((\Delta \phi)^4)\Big ].
\label{flow-approx}
\end{eqnarray}
The procedure for obtaining the number of jet particles and the jet
opening angle from the data are now clear. After the computation of the
flow parameters $v_m$'s, one has to fit $v_m^2$ with a polynomial in
$m^2$ as shown in the eq(\ref{flow-approx}) above. Then the
intercept on y-axis gives $N_j$ ( note that $N$ is known from the
data ) and the coefficient of $m^2$ term  is related to $\Delta
\phi$. 

Although the approximate equation for the flow ( eq(\ref{flow-approx}))
has been derived for uniform distribution, it can be generalized for 
arbitrary probability distribution of jet particles, provided that the 
probability function $P_2(\phi)$ is
sharply peaked at some angle $\phi_0$. In such a case, we can choose 
the x axis along $\phi_0$ ( as before ) and expand the cosine function
in powers of ($\phi_1 - \phi_2$). Keeping up to quadratic terms in the 
expansion, we get
\begin{eqnarray}
v_m^2 & =&  \int d\phi_1 d\phi_2 
\cos(m(\phi_1 - \phi_2)) P_2(\phi_1) P_2(\phi_2)   \nonumber \\
&=& \int d\phi_1 d\phi_2 \Big[ 1 - \frac{m^2(\phi_1 - \phi_2)^2}{2}
\Big] P_2(\phi_1) P_2(\phi_2) \nonumber \\
&=& \frac{N_J^2}{N^2}\Big [ 1 - m^2 \sigma^2 \Big ].
\label{var}
\end{eqnarray}
Here $\sigma = \sqrt{<\phi^2> - <\phi> ^2}$ is the variance of the 
distribution  $P_2(\phi)$. Note that the variance $\sigma$ is related to 
the spread of the jet particles in azimuthal angle. Particularly for 
uniform distribution defined in 
eq(\ref{dist_fun}) above, $\sigma^2 = \Delta \phi ^2 / 12$. Thus, for 
any sharply peaked probability distribution functions of jet particles, 
we can plot $v_m^2$ {\it vs} $m^2$ and from a linear fit to the curve 
one would get the number of jet particles and the opening angle of the
jet.

From the expressions of the flow coefficients eq(\ref{flow},
\ref{flow-approx} ) it is clear that $v_m^2$ would have sufficiently 
large and measurable if the number of jet particles constitute a 
sufficiently large fraction of the total number of particles $N$. 
Generally, this will not be the case. However, the jet particles 
are expected to have large transverse momentum and the background
particles have exponentially falling transverse momentum distribution.
One can therefore use transverse momentum cut to eliminate a large 
fraction of background particles and 
enhance the fraction of jet particles. This is expected to improve the 
determination of the jet 
parameters. We have performed the calculations with different transverse 
momentum cuts and these results will be discussed in next the section.   

\subsection{Two Jets}
Let us now consider a case when there are two jets in the given rapidity
window. We shall assume that the number of jet particles are $N_{J_1}$ 
and $N_{J_2}$ and the respective opening angles are $\Delta \phi_1$ and 
and $\Delta \phi_2$. Without loss of generality, we can choose the jet 
angle of the first jet to be zero and that of the second jet to be 
$\Phi$. Then the probability distribution function for this case is
\begin{eqnarray}
P(\phi) & = & P_1(\phi) + P_2(\phi) + P_3(\phi)
\label{dist_fun2}
\end{eqnarray}
with
\begin{eqnarray}
P_1(\phi) & = &  \frac{N_b}{2 \pi N} ~~~~~~~~ 0 < \phi < 2 \pi \nonumber \\
P_2(\phi) & = &  \frac{N_{J_1}}{\Delta \phi_1 N} ~~~~~~~~
 -\frac{\Delta \phi_1}{2} < \phi <  \frac{\Delta \phi_1}{2} ,
\nonumber \\
& = & 0  ~~~~~~~~~~~~~~~~~~ {\rm otherwise} \nonumber \\
P_3(\phi) & = &  \frac{N_{J_2}}{\Delta \phi_2 N} ~~~~~~~~
 \Phi-\frac{\Delta \phi_2}{2} < \phi <  \Phi+\frac{\Delta \phi_2}{2} ,
\nonumber \\
& = & 0  ~~~~~~~~~~~~~~~~~~ {\rm otherwise}. \nonumber
\end{eqnarray}
As in the previous case, one can evaluate the flow coefficients 
analytically. As before, the background particles do not contribute. The 
result is, 
\begin{eqnarray}
  v_{m-2J}^2 &=& \Big [j_0 \Big ( \frac {m \Delta \phi_1} {2} \Big )
\Big ]^2 \frac{N_{J_1}^2}{N^2} 
 + \Big [j_0 \Big ( \frac {m \Delta \phi_2} {2} \Big )\Big ]^2
\frac{N_{J_2}^2}{N^2} \nonumber \\
 & & + \Big [ j_0 \Big ( \frac {m \Delta \phi_1} {2} \Big )
j_0 \Big ( \frac {m \Delta \phi_2} {2} \Big )
     \cos(m\Phi) \Big]
\Big[\frac{2N_{J_1}N_{J_2}} {N^2}\Big ]
\label{gen-2j}
\end{eqnarray}
The meaning of the three terms in the equation above are obvious. The 
first two terms arise when both of the particles are from jet 1 and jet 2
respectively and the last term arises from the case when one particle is 
from jet 1 and the other is from jet 2. As mentioned earlier, when one 
of the particles is from background, the contribution to the flow
coefficient vanishes.
 
Consider a special case when the opening angels of jet one with
 $N_{J_1}$ particles and jet 
two with $N_{J_2}$ particles are same. Then eq(\ref{gen-2j}) reduces to
\begin{eqnarray}
  v_{m-2J}^2 =
\Big [j_0 \Big ( \frac {m \Delta \phi} {2} \Big )\Big ]^2
    \Big [ \cos(m\Phi) \Big(\frac{2 N_{J_1}N_{J_2}} {N^2}\Big ) + 
\Big( \frac{N_{J_1}^2+N_{J_2}^2}{N^2} \Big )\Big ].
\label{gen1-2j}
\end{eqnarray}
When $\Phi$, the angle between two jets is zero, the result in 
eq(\ref{gen1-2j}) reduces to the single jet result (eq(\ref{flow}) with 
the number of jet particles $N_{J_1} + N_{J_2}$. A case of particular 
interest is when $\Phi$ is equal to $\pi$. For that case, $\cos(m\Phi)$
is $(-1)^m$ and therefore the flow coefficients for odd values of m 
are much smaller than those for even values. In fact, if the number of 
particles in each jet are equal, the flow coefficients for odd m would 
vanish theoretically. Since a pair of jets are produced from a hard 
collision of two partons from two nuclei, $\phi$ for the two jets will 
be close to $\pi$\cite{jet_angle}.  However, if there is substantial 
jet quenching, the leading partons will loose some momentum to particles
in quark-gluon plasma. In that case, the angle $\Phi$ will differ from 
$\pi$. Thus, in the case of two jets, the behavior of $v_m^2$ as a 
function of $m$ will be of interest.

In general, there are five parameters to be determined ( number of
particles in two jets, two opening angles and the angle between two
jets ). Since one can measure $v_m$'s for $m$ going from 1 to 5 or 6 ( at
the most ), we do not expect that these parameters can be determined to 
sufficient accuracy. But, probably the most important quantity in case
of two jet events is the angle between two jets and that can probably 
be inferred.

\subsection{More Than Two Jets}

Our theoretical analysis can be extended to events having
more than two jets. We do not do this for the following reasons.
\begin{enumerate}
\item 
Unlike two jet events, we do not expect a possible
correlation in the angles between the jets. In case of two jet
events, momentum conservation implies that the angle between 
two jets is close to $\pi$. As mentioned in the preceding
subsection, this produces definite pattern in $v_m^2$'s. In
case of events with three and more jets, no such pattern is 
expected.
\item 
In nucleus-nucleus collisions, more than two jets will be produced
when

a) a single nucleon-nucleon collision produces more than two jets

b) or when two uncorrelated nucleon-nucleon collisions produced jets.

The probability of latter is very small since, as estimated in 
introduction, the chance of (hard) jet production itself is small.
In case of the former, the multiple jet production probability would
be related to $\alpha_s$, the strong coupling constant. The 
estimates indicate that the ratio of number of events
with three or more jets to those with two jets is about 0.1 when the
jets have large transverse momentum\cite{cdf95}.
\item
In order to control the number of "background" particles, it is
better to do our analysis in a restricted rapidity window. The jet
particles are expected to have a rapidity distribution in $\Delta\eta 
\sim 0.5$ and therefore the
$\eta$-window for the analysis should be between 0.5 and 1.0. 
Because of this, it is likely that some of the jets in a multi-jet
events will lie outside this window. This would further reduce the
possibility of encountering more number of jets during the analysis.
\item
Further, we do not think that
it would be possible to extract meaningful information about
the jets in multi-jet events because the number of parameters in
this case are large (jet angle and width for each jet and relative angles 
between jets) and the number of flow coefficients which can be measured 
reliably is restricted to five or six (at the most).
\end{enumerate}

It is however pertinent to ask if the flow analysis gives some 
information regarding the multi-jet events. In particular,
one would like to know whether one would be able to classify
the events into no jets, one jet, two jet and multi-jet events
after the flow analysis. We believe that this may be possible
with some level of confidence. For example, the events without
a jet, which would constitute the bulk of the events, would yield
the flow coefficients similar to those obtained for uniform random
distribution of particles (in azimuthal angle). For single
jet events, the flow coefficients $v_m^2$'s will be larger and
decreasing with $m$. For two jet events, these would be oscillating
and if the jet angle is close to $\pi$, the coefficients for odd
values of $m$ would be small. For events with more than two jets,
we expect the particle distribution to be closer to the random 
distribution.

\section{The Calculations And Discussions}

The background particles in an event are constructed by taking the 
output from an event generator. For this purpose we have used the 
generator\cite{simul} which uses dynamical hadron string cascade model. 
For our analysis, we have considered 130 GeV collision of gold nuclei. 
To these `background' particles, the jet particles are added by hand. We 
choose different numbers of jet particles and assume that their 
transverse momenta are larger than 2 GeV. The direction of the jet 
as well as the azimuthal angle of the jet particles are chosen randomly. 
We, however assume that the probability of the azimuthal angle of the 
jet particles is constant over the opening angle $\Delta \phi$. In the 
following we shall consider the analysis when the event does not have a 
jet, there is only a jet ( no background ) and both jet and background 
particles are present. 

\subsection{Events Without Jet} 

Let us first consider the case where the event consists of only the 
background particles and no jet. Theoretically, the values
of the flow coefficients are all zero as the background particles are 
uniformly distributed. However the measured coefficients differ from zero
because of the fluctuations. Fig(\ref{nojet}) shows the plot of the 
square of the flow coefficients for one of the events. The figure shows 
that the flow coefficients are non-zero and sometimes negative. This 
clearly indicates that the observed flow coefficients are indeed due to 
fluctuations. Further more, the values of the flow coefficients are 
smaller for lower transverse momentum cutoff. This is because the number of 
particles, in case of the lower cutoff are larger and therefore the fluctuations 
are smaller. 

\begin{figure}[ht]
\epsfxsize=7cm
\begin{center}
\includegraphics{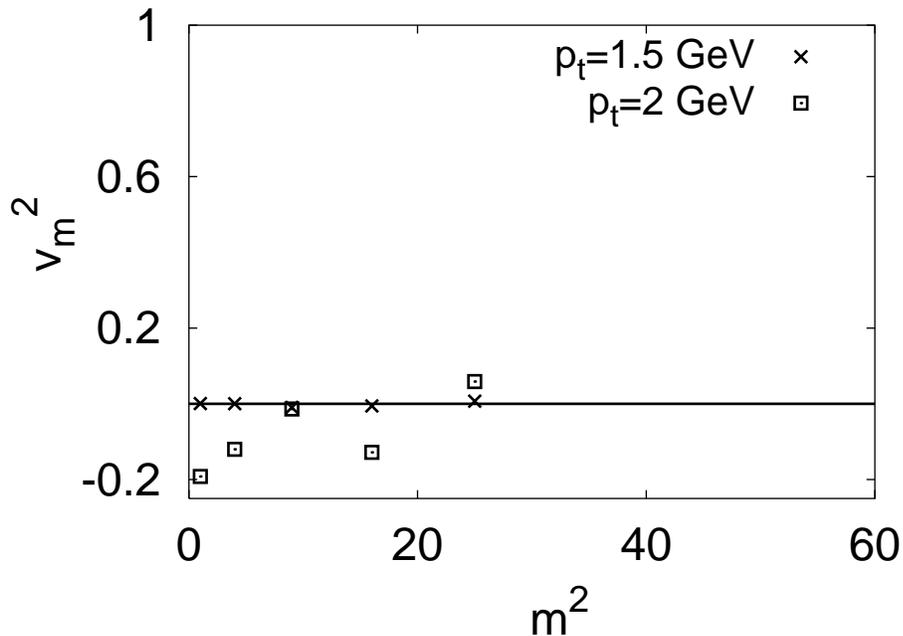}
\end{center}
\caption{Plot of $v_m^2$ {\it vs} $m^2$ for the case of no jet. Note that the 
values of $v_m^2$'s are close to zero for $P_t$ cut of 1.5 GeV. Also 
note that sometimes the extracted values of $v_m^2$ are negative, 
clearly indicating that these are essentially due to fluctuations.}
\label{nojet}
\end{figure}

The computed  values of the flow coefficients for the events without 
a jet set a lower limit on the values of the flow coefficients from 
which meaningful information about jet characteristics can be 
determined. In particular, one may note that for smaller transverse 
momentum cutoff, the flow coefficients in absence of a jet are closer
to zero as there are lot more background particles. So, in this case,
in order to get reliable information, the flow coefficients in the 
presence of a jet need to be larger than the flow coefficients shown 
here.  

\subsection{Jet Only Events}

Let us now consider the case where there are no background particles. 
This, in a way,  is an ideal case and we expect to obtain the best 
results from our analysis. Here we display the results for the extracted 
values of number of jet particles ( $N_o$ ) and the opening angle 
($\Delta \phi_o$) and compare these with the corresponding input values 
( $N_g$ and $\Delta \phi_g$ respectively ) in Fig(\ref{jetonly}). The 
calculations are performed for different values of number of particles 
and opening angle of the jet. Also, the calculations are repeated a 
large number of times with different sets of background particles as well 
as jet particles so as to estimate the possible error in the extracted 
quantities. Fig(\ref{jetonly}) shows that the extracted values of the 
number of jet particles are very close to the input values. Also, the 
extracted values of the opening angle are close 
to the input values for $\Delta \phi$ smaller than 0.75. Note that the 
expected opening angle of a jet is about this order. These results seem 
to indicate that it may be possible to determine the number of jet 
particles reasonably well. As for opening angles, we find that the 
determination may 
not be very much reliable if the opening angle exceeds 0.75.
 
\begin{figure}[ht]
\epsfxsize=8cm
\begin{minipage}{0.4\textwidth}
\epsfbox{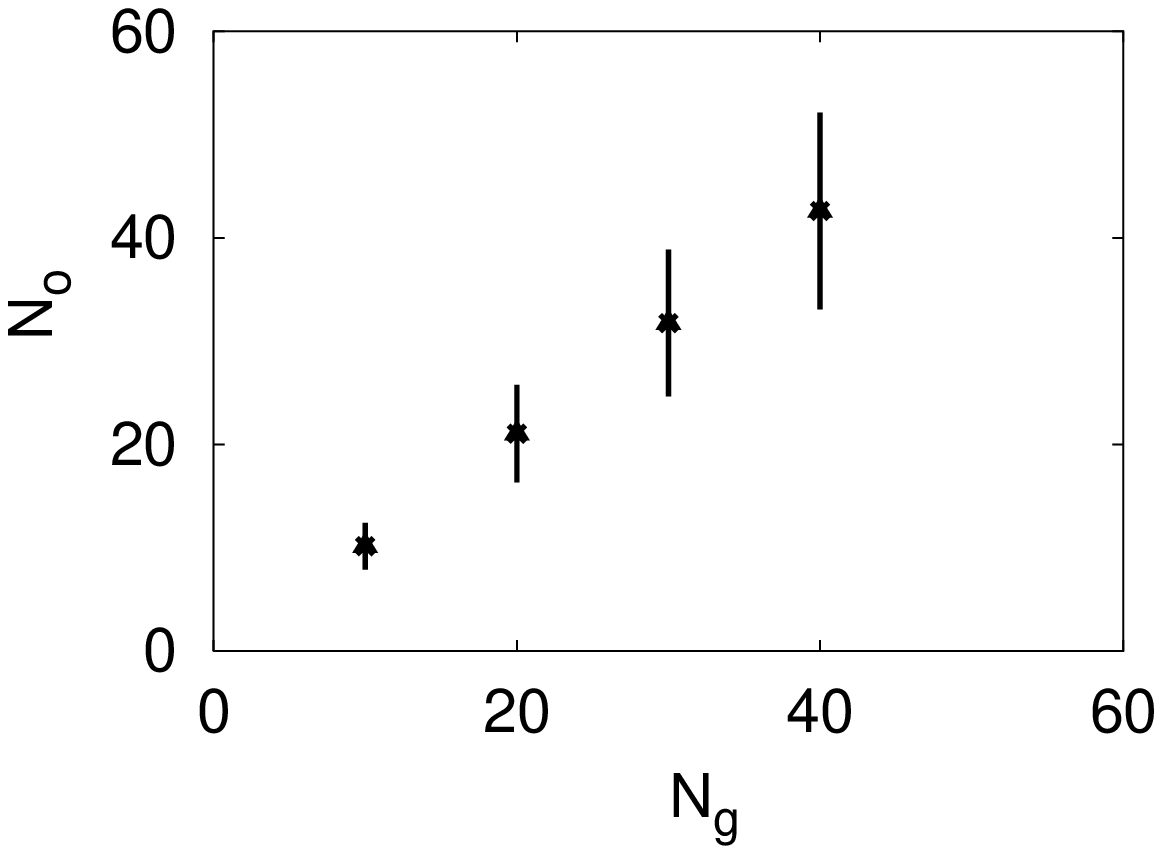}
\end{minipage} \
\begin{minipage}{0.4\textwidth}
\hspace*{1cm}
\epsfbox{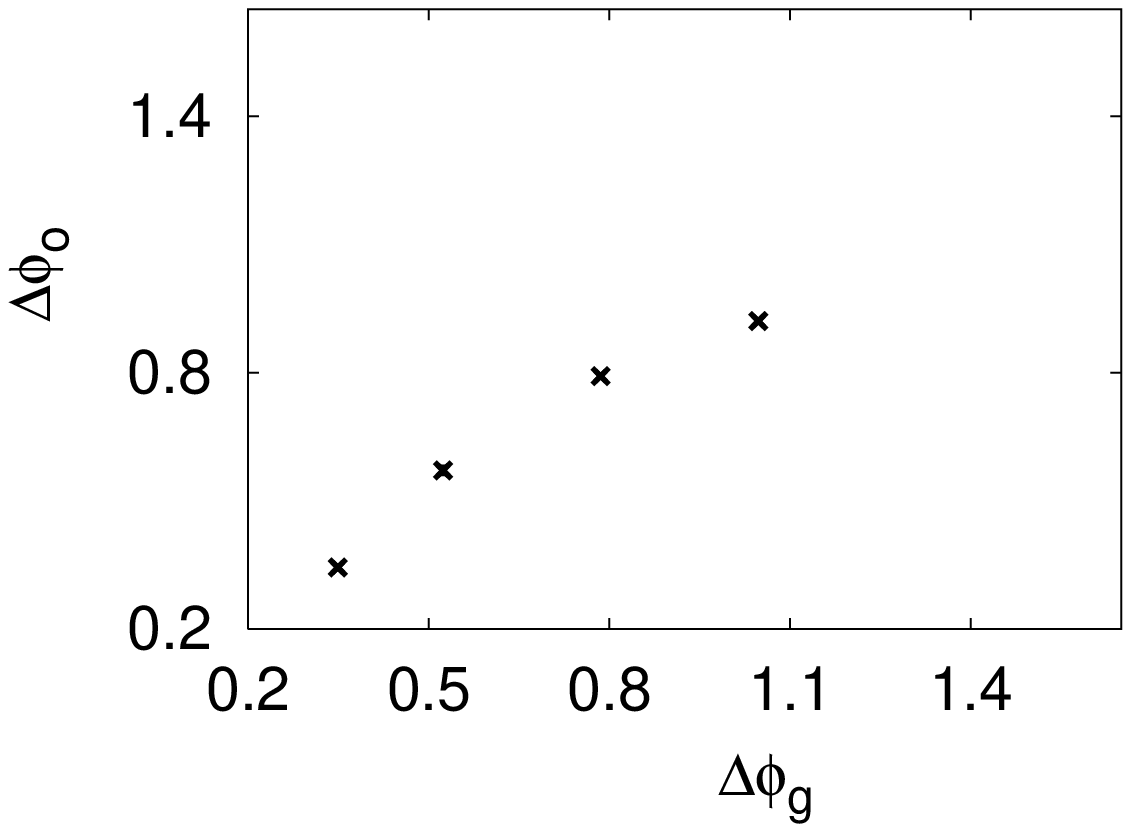}
\end{minipage}
\caption{Plot of $v_m^2$ {\it vs} $m^2$ for the case of no jet.}\label{jetonly}
\end{figure}

\subsection{Events With Jet and Background}
We shall now consider the analysis of events in which a jet is produced 
in the presence of  `background' particles. For this analysis, a large 
number of events with impact parameters 1 fm and 6 fm ( corresponding to 
the central and peripheral events ) have been considered. The reason for 
choosing these two types of events is as follows. The number 
particles produced in a central event are large and this may make the 
determination of jet difficult. But such events are expected to have 
negligible elliptic flow. That is, the flow coefficients ( including 
the elliptic flow ) for the background particles are expected 
to be close to zero. This, on the other hand, is likely to help the jet 
characterization analysis. For the peripheral events, on the other hand, 
the number of background particles is smaller but the event are likely 
to have nonzero elliptic flow. The peripheral events are of interest for
one more reason. For these events, the distance traveled by the leading 
parton in the jet is expected to be different for in-plane and 
out-of-plane direction of the parton. This would mean that 
the properties of jets in these two directions would be different if jet
quenching is important.

It is clear that, for our method to work, the number of background 
particles should not be very large, so that the jet is not completely 
swamped  by the background particles. One way of controlling the 
number of background particles is to remove the particles having 
transverse momenta smaller than certain value. This would remove 
background particles but not the jet particles if the cutoff momentum is 
sufficiently small. In fact, a very large fraction of the 
background particles have transverse momenta smaller than 1 GeV. So, a 
transverse momentum cutoff larger than 1 GeV would help in the analysis. 
On the other hand, the cutoff should not be too large since this may 
remove some of the jet particles also.  Further more, when the number of 
background particles is very small, the fluctuations in the flow 
coefficients due to these background 
particles are large and this would affect the determination of the jet 
characteristics. We have therefore analyzed the events by including the 
background particles having transverse momenta larger than 1.5 and  2 GeV. 

\begin{figure}[ht]
\epsfxsize=8cm
\begin{minipage}{0.4\textwidth}
\epsfbox{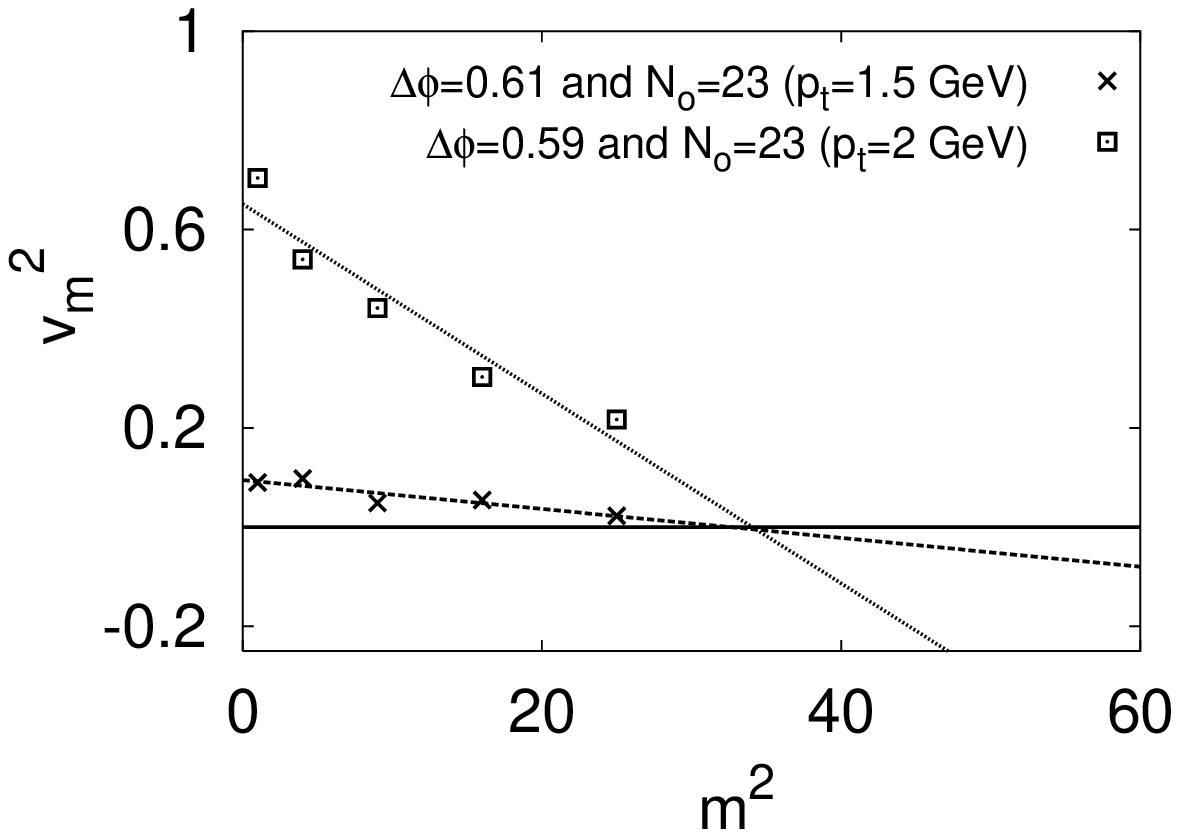}
\end{minipage} \
\begin{minipage}{0.4\textwidth}
\hspace{1cm}
\epsfbox{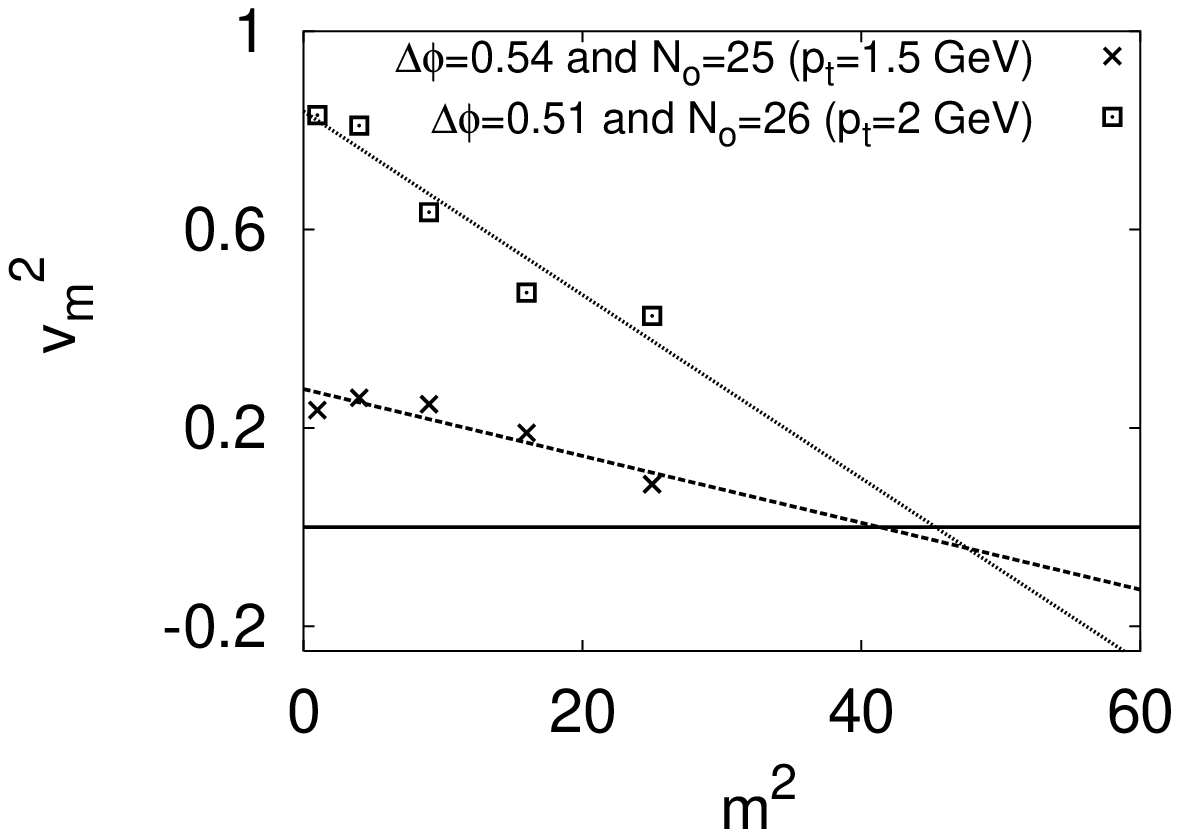}
\end{minipage}
\caption{Plot of $v_m^2$ {\it vs} $m^2$. The left panel is for the central (N=74
and 28 for $p_t$=1.5 and 2 GeV, respectively) and the right panel is for the
peripheral (N=48 and 28 for $p_t$=1.5 and 2 GeV, respectively) collisions.
}
\label{jetplus}
\end{figure}

Fig(\ref{jetplus}) displays a plot of the square of the flow 
coefficients $v_m$ {\it vs} $m^2$ for the two transverse momentum cuts. The 
opening angle for this graph is $\pi/6$ and the number of jet particles 
is 20. The left panel is for the central collision and the right panel 
is for the peripheral collision. Note that the computed flow coefficients
are significantly larger than those for no jet ( see Fig(\ref{nojet} ). 
Thus we expect that the extracted values of jet parameters 
are reliably determined. The extracted values of the number of jet 
particles and the opening angle are shown in the figure. We find that 
the extracted values of number of jet particles as well as the opening 
angle are within 25 \% of the input values.

The plots in Fig(\ref{jetplus}) are for the analysis of one particular 
event. We have repeated the analysis for a large number of such events.
As expected, the extracted values fluctuate and from the fluctuations 
we have determined the errors in the extracted quantities. These are 
displayed in Fig(\ref{numbers}).

\begin{figure}[ht]
\begin{minipage}{0.4\textwidth}
\epsfxsize=8cm
\epsfbox{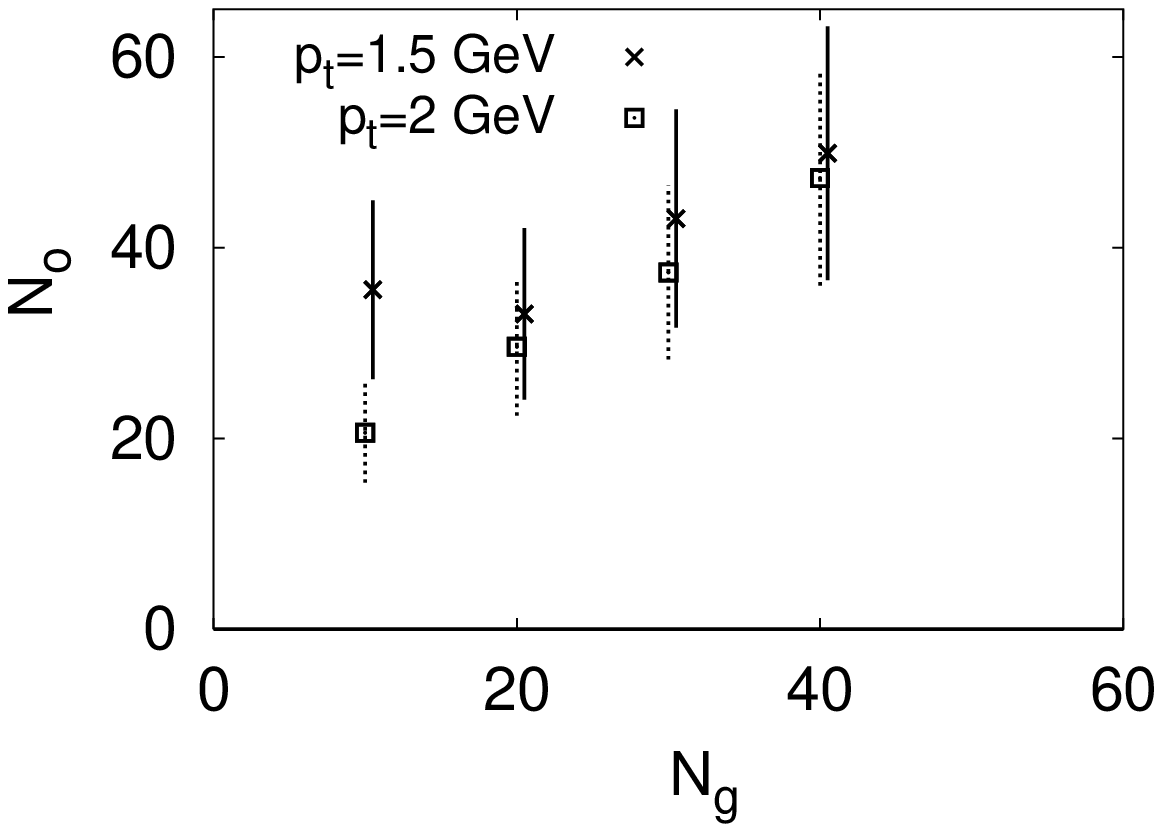}
\epsfxsize=8cm
\epsfbox{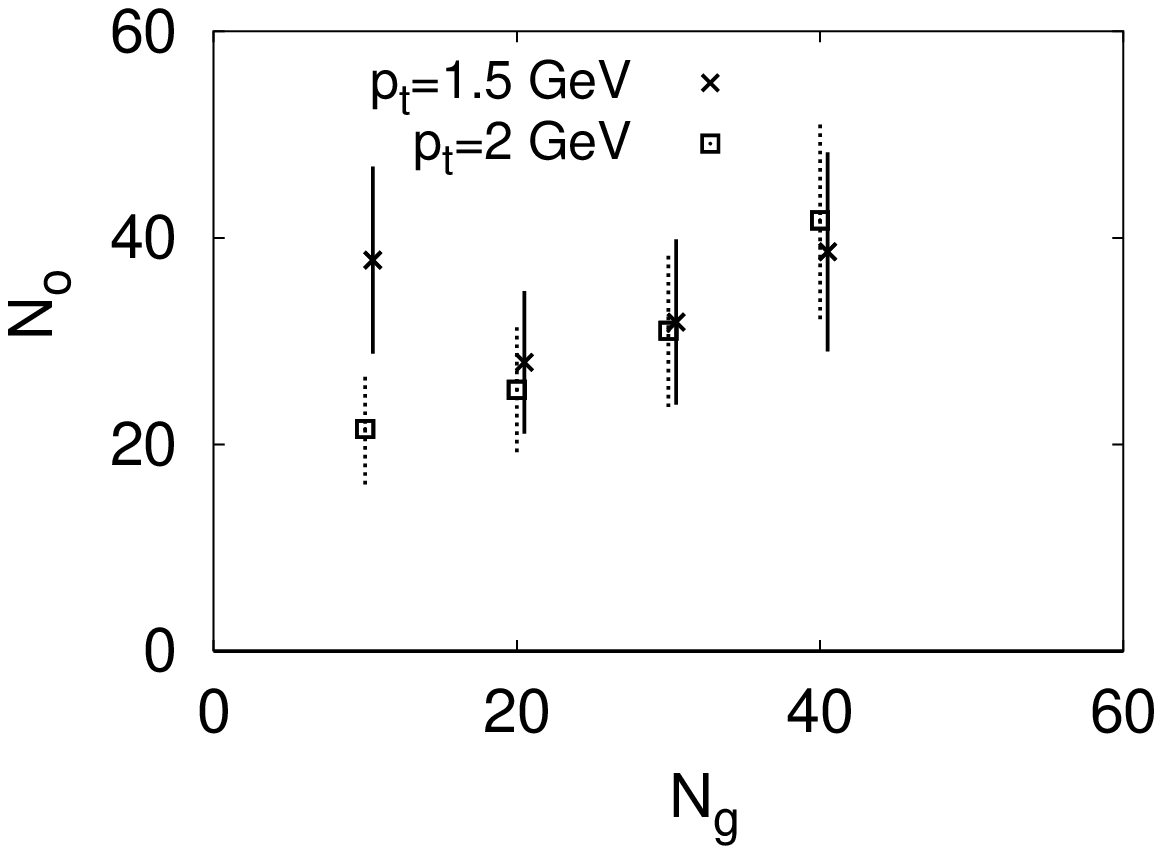} 
\end{minipage} \ 
\hspace*{1cm}
\begin{minipage}{0.4\textwidth}
\epsfxsize=8cm
\epsfbox{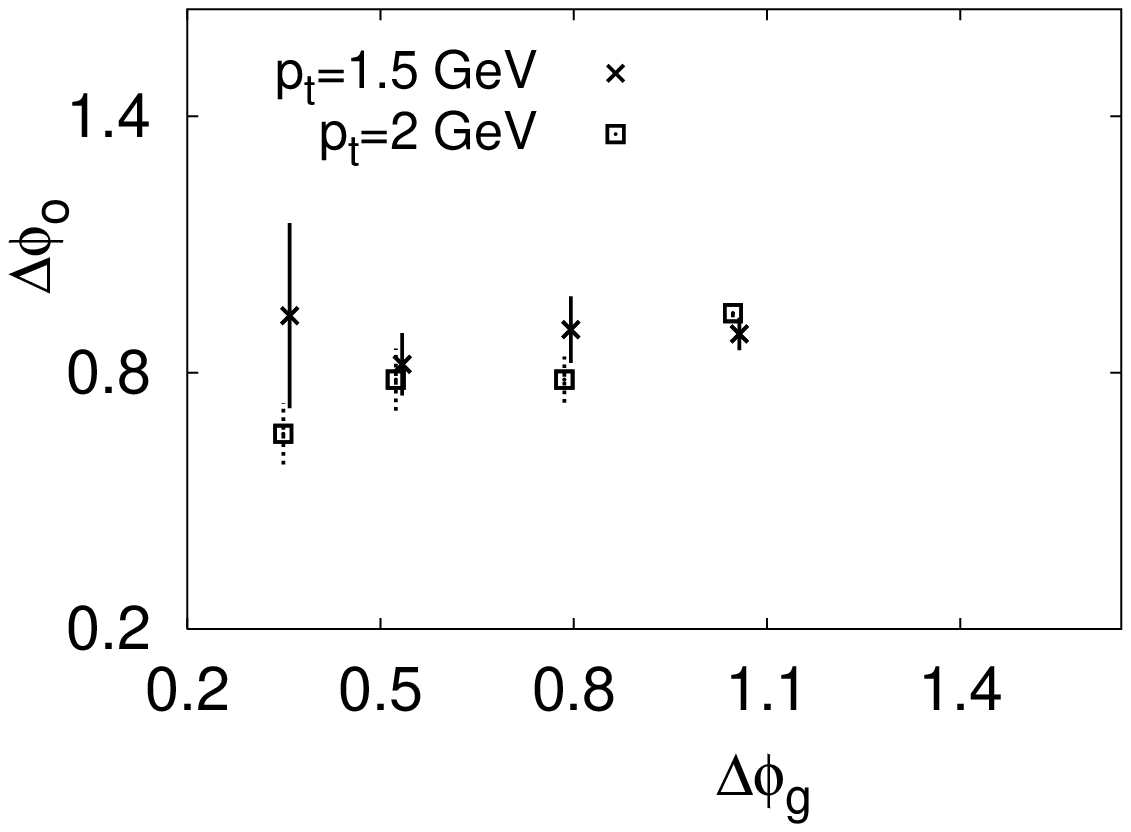}
\epsfxsize=8cm
\epsfbox{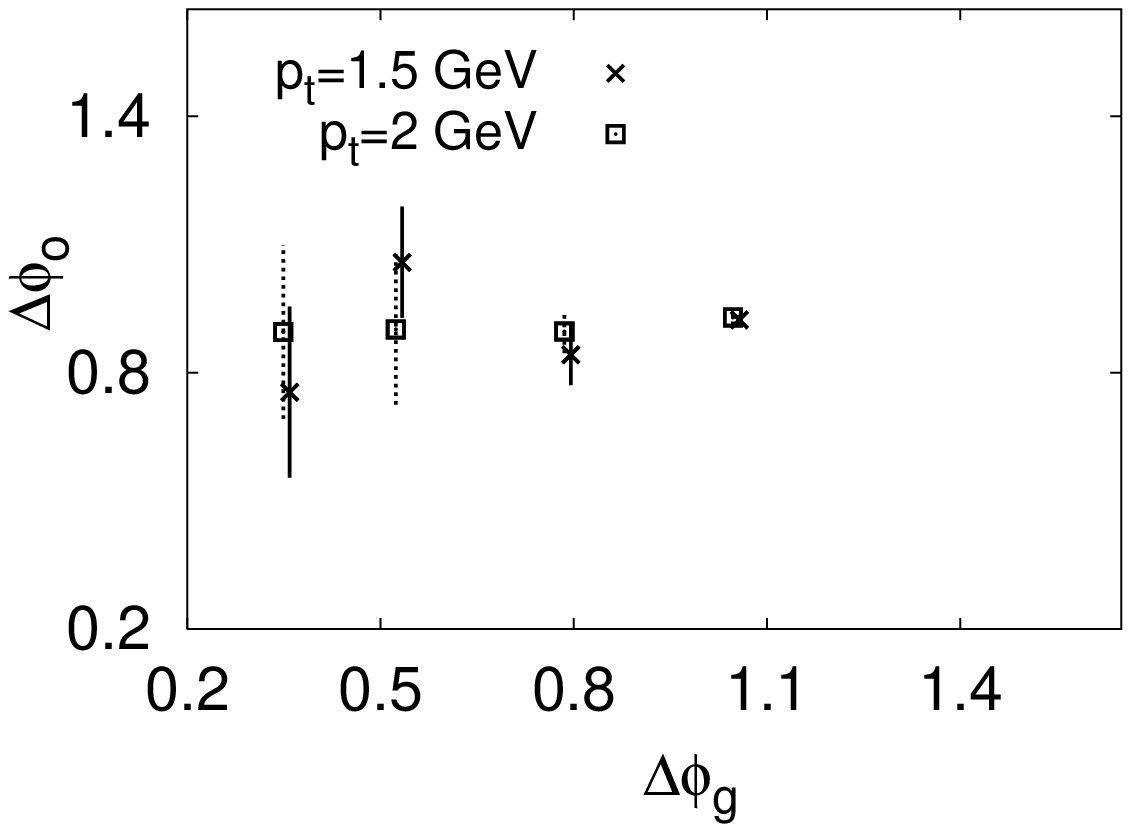}  
\end{minipage}
\caption{Plot of the extracted numbers of jet particles (opening angles) 
{\it vs} the corresponding input values are shown in left(right) panels. 
The top (bottom) panels are
for the central (peripheral) collisions. The data points and error bars 
for $p_t=1.5$ GeV have been shifted marginally 
to distinguish from $p_t=2$ GeV.}
\label{numbers}
\end{figure}

The plot of the number of jet particles extracted from the analysis {\it vs }
the corresponding input value is shown in the left panels of 
Fig(\ref{numbers}). The top panel is for central collisions and the 
bottom panel is for peripheral collisions. The results are for the 
opening angle of $\pi/6$. 
The plot shows that there is a linear correlation between the particle
numbers extracted and the corresponding input numbers when the number 
of jet particles are larger than 10. Also, the correlation is better for 
2 GeV cutoff. We find that the results for central and peripheral 
collisions are similar.

The right panels of Fig(\ref{numbers}) displays the extracted values of 
the opening angle {\it vs} the input opening angle. The results are for 
twenty 
jet particles. This figure shows that there is a correlation between 
extracted and input opening angles but it is not as strong as that for 
the 
number of jet particles. Particularly, the extracted opening angle does 
not increase linearly for larger input opening angles. One reason for 
this is that for larger angles, the expansion in the powers of $m^2$ 
fails for large $m$ and these results in the saturation of the computed 
opening angle. 

The extracted values of the number of jet particles for different input
opening angles have similar behavior as that observed in left panels
of Fig(\ref{numbers}).  Thus, it seems that the extraction of the 
number of jet particles is quite robust but that of the jet opening 
angle from the flow analysis is less reliable.

We would like to note here that, although the extracted flow coefficients 
are smaller for 1.5 GeV cutoff, one is still able to obtain some information 
about jets. This is because the extracted flow coefficients are still 
sufficiently larger than those for the background. When the cutoff is reduced 
below 1 GeV, we find that the extracted coefficients are extremely small. 
That is, the background particles essentially swamp the jet particles. In such
cases a meaningful extraction of jet properties is not possible. We 
find that, in order to obtain significant jet information, the number of jet 
particles should constitute more than 5 \% of the total number of particles. 
It should be apparent that this requirement reduces when the jet angle is
smaller.

\subsection{Two Jets plus Background}

\begin{table}
\caption{particles from two jets, $v_{m-2J}^2$ {\it vs} $m$ for different 
parameters.}
\vskip 0.1 in
\begin{center}
\begin{tabular}{cccccccccccc}
\hline
\multicolumn{1}{c}{$p_t$(GeV)}&
\multicolumn{1}{c}{$\phi$}&
\multicolumn{1}{c}{$\Delta\phi$}&
\multicolumn{1}{c}{$N_{J1}$} &
\multicolumn{1}{c}{$N_{J2}$} &
\multicolumn{1}{c}{$v_{1-2J}^2$} &
\multicolumn{1}{c}{$v_{2-2J}^2$} &
\multicolumn{1}{c}{$v_{3-2J}^2$} &
\multicolumn{1}{c}{$v_{4-2J}^2$} \\
\multicolumn{1}{c}{$$}&
\multicolumn{1}{c}{$$}&
\multicolumn{1}{c}{$$}&
\multicolumn{1}{c}{$$} &
\multicolumn{1}{c}{$$} &
\multicolumn{1}{c}{$$} &
\multicolumn{1}{c}{$$} &
\multicolumn{1}{c}{$$} &
\multicolumn{1}{c}{$$} \\
\hline
$1.5$&$\pi$ &$\frac{\pi}{6}$&10&20&0.00047&0.14&0.0091&0.196\\
$1.5$&$\pi$ &$\frac{\pi}{6}$&20&20&0.0069&0.18&0.00092&0.236\\
$2.0$&$\pi$ &$\frac{\pi}{6}$&10&20&0.096&0.74&0.123&0.597\\
$2.0$&$\pi$ &$\frac{\pi}{6}$&20&20&0.0008&0.78&0.0072&0.633\\
\hline
$2.0$&$\frac{2 \pi}{3}$ &$\frac{\pi}{6}$&10&20&0.265&0.206&0.519&0.198\\
$2.0$&$\frac{2 \pi}{3}$ &$\frac{\pi}{6}$&20&20&0.192&0.195&0.586&0.139\\
$2.0$&$\frac{5 \pi}{6} $ &$\frac{\pi}{6}$&10&20&0.146&0.543&0.410&0.163\\
$2.0$&$\frac{5 \pi}{6} $ &$\frac{\pi}{6}$&20&20&0.052&0.587&0.326&0.178\\
\hline
$2.0$&$\pi$ &$\frac{\pi}{9}$&10&20&0.097&0.779&0.132&0.746\\
$2.0$&$\pi$ &$\frac{\pi}{9}$&20&20&0.0007&0.819&0.0063&0.776\\
jets only&$\pi$ &$\frac{\pi}{6}$&10&20&0.157&0.915&0.135&0.696\\
jets only &$\pi$ &$\frac{\pi}{6}$&20&20&0.0025&0.921&0.0033&0.715\\
\hline
\end{tabular}
\end{center}
\end{table}

The computed flow coeficients when two jets are present are displayed 
in Table I(columns 6 through 9). The first three columns give transverse
momentum cut, angle between the two jets ( $\phi$ ) and opening angles
of the jets. The next two columns give the number of particles in the two
jets. The angle $\phi$ between the two jets has ben chosen 
to be larger than $120 \;^\circ$. As discussed in the previous section, 
we expect that the angle between two jets is close to $\pi$. The opening 
angles of both the jets have been chosen to be equal.

We find that the determination of individual jet parameters from the 
flow analysis is not 
possible because the number of measured flow coefficients $v_m$'s does 
not exceed 6. Therefore a detailed analysis, like in the case of a 
single jet, cannot be done. Never the less, the table clearly shows that
the flow coeficients for odd $m$'s are generally smaller than those of 
even $m$'s, particularly when the angle between the jets is $\pi$. It 
may be noted that this odd-even effect is noticable even when the number 
of particles in each jet differs significantly. But the effect reduces 
considerably when the angle is reduced to $150 \;^\circ$ or 
less\cite{phi}. 
So, it seems that a good estimate of the angle between the two jets from
the flow analysis may possible.

\section{Summary}

A method of extracting jet characteristics from flow analysis is 
discussed in this work. We have shown that when the data has a jet with
the number of jet particles in excess of 10, the higher flow 
coefficients are large enough to be measured. In particular, we show 
that transverse momentum cuts can be used to enhance the effect of jet
particles on the flow coeficients. Further, by analyzing the flow 
coefficients of different orders, we show that it is possible to 
extract the information of jet characteristics, namely the number of 
particles in the jet and jet opening angle. We believe that this is an 
interesting result and can be used to analyze the heavy ion collision 
data and extract the jet information. Our calculations show that in 
order 
to extract meaningful results, it is necessary to use transverse 
momentum cut to reduce the background. We show that a cut between 1.5 
and 2 GeV works well. This is nice since the jet particles are expected 
to have transverse momenta larger than these values. 

Although we do not demonstrate, we believe that instead of the 
transverse momentum cuts, one may compute the transverse 
momentum ( or transverse energy ) weighted flow coefficients
\begin{eqnarray}
  \label{eq:2}
  b^{p_T}_m(p_0) & = &\frac{\sum_{n} p_T(n) \cos(m \phi(n) )}{\sum_{n}} 
	\nonumber \\
  & = & \frac{1}{N(p_0)} \sum_{n, p_T(j) > p_0} p_T(n) \cos(m \phi(j) ).
\end{eqnarray}
One may even use higher powers of the transverse momentum in 
eqn(\ref{eq:2}) above. Actually, such flow coefficients do de-emphasize 
the particles having smaller momenta, which are predominantly the 
background particles. A bonus, in such calculations is that such flow 
coefficients will give information on the transverse momentum or energy 
carried by the jet particles.  

We have also considered the case when two jets are present in the given 
rapidity window. For such a case, we have argued that an important 
information to be determined in this case is the angle between the 
two jets. We have shown that it may be possible to estimate this angle
between the two jets.
%Added

The method developed in the present work is not useful to obtain the 
information of more than three jets. This is because the number of 
parameters in such a case are larger than the number of flow coefficients
which can be measured meaningfully. However, the multi-jet events are
expected to yield the flow coefficients which are similar to those
obtained for the background particles.

In the present work, the jet particles have been introduced by
hand. This has been done because, at this stage, we want to analyse
jet events with known jet properties (number of particles and 
opening angles) so that
we can determine how faithfully these are extracted from the analysis.
The results are encouraging when one has a single jet. The computations
also indicate that two-jet events are also identified if the angle
between two jets is close to $\pi$. The next step is to use our 
procedure on the data created using generators which include jet production 
algorithm\cite{prog}. If this succeeds, one can think of applying the 
procedure to RHIC data.

\section{Acknowledgements} 
One of us (PKS) would like to thank A. Ohnishi and Y. Nara for helpful
discussions.


\begin{thebibliography}{99}
\bibitem{jet_angle} For the jets produced in $e^+-e^-$ collisions, the
angle between the two jets is $\pi$. In case of the jets produced in
nucleus-nucleus collisions, the angle will be slightly different from 
$\pi$ as the two partons will have some transverse momentum before the
hard collision. However, the deviation is expected to be small as the
parton transverse momentum before the collision is not very large.
\bibitem{phi} It is interesting to note that when $\phi$ the angle 
between two jets is $120 ^\circ$, $\cos 3\phi$ is $+1$ and $v_3$ is 
largest.
\end{thebibliography}
\end{document}